# Life Before Fermi – Back to the Solar System

Dr David L Clements

Astrophysics Group, Blackett Lab, Physics Department, Imperial College, Prince Consort Road, London SW7 2AZ, UK

**Abstract**

The existence of intelligent, interstellar traveling and colonising life is a key assumption behind the Fermi Paradox. Until recently, detecting signs of life elsewhere has been so technically challenging as to seem almost impossible. However, new observational insights and other developments mean that signs of life elsewhere might realistically be uncovered in the next decade or two. We here review what are believed to be the basic requirements for life, the history of life on Earth, and then apply this knowledge to potential sites for life in our own Solar System. We conclude that the necessities of life – liquid water and sources of energy – are in fact quite common in the Solar System, but most potential sites are beneath the icy surfaces of gas giant moons. If this is the case elsewhere in the Galaxy, life may be quite common but, even if intelligence develops, is essentially sealed in a finite environment, unable to communicate with the outside world.

1. Introduction

The central realisation of the Fermi Paradox (see eg. Brin, 1983; Hart, 1975) is that a space-faring civilisation, spreading from habitable planet to habitable planet even at speeds substantially below that of light, can visit every star in the Galaxy on a timescale of 50-100Myrs. This is a very small timescale compared to the age of the Galaxy (~10 Gyrs) or the age of the Sun (4.5 Gyrs). Our hypothetical aliens should thus already be here, and yet they are not. This can be used as an argument against the existence of intelligent extraterrestrials (eg. Tipler, 1980, Sandberg et al., 2018), but our own existence is proof that intelligent life can and does arise in the Galaxy. This is the central puzzle of the Fermi Paradox.

Many solutions have been suggested to the Fermi paradox (eg. Brin, 1983; Webb, 2002, Sandberg et al., 2017) but all this discussion has taken place inside a vacuum of observational evidence. This may change in the next decades with the advent of instruments like the Square Kilometre Array (SKA), which is sufficiently sensitive that it should be able to detect air traffic control radars out to a distance of 10pc with relative ease (Siemion et al., 2015). However, there are other steps on the road to the Fermi Paradox that we can examine with observational evidence from our own Solar System that is already available. We aim to do this in the current paper through looking first at the history of life on Earth, then at the resources that seem to be essential for the emergence of life, and then an examination of the sites in the Solar System where these resources may exist.

The rest of the paper is structured as follows. In Section 2 we look at the history of life on Earth, incorporating the latest information about the emergence of both life, habitability, and currently habitable environments. In Section 3 we look at the resources that appear to be essential for the existence of life. Then in Section 4 we examine which bodies in the Solar

System, on the basis of available resources, might be capable of hosting life. We discuss these results and draw our conclusions in Section 5.

2. **The Early History of Life on Earth**

Determining the date at which life emerged on Earth is a central goal of paleobiology and a very active field. In recent years, the date of the earliest known life on Earth has been pushed to ever earlier epochs. Fossilised stromatolites – distinctive sedimentary rock formations produced by layers of photosynthesising bacteria – and microfossils clearly show that life was well established 3.5 Gyr ago (Schopf, 2006). More recent studies have pushed back the date of stromatolites to 3.7 Gyr ago (Nutman et al., 2016), just 800 Myr after the formation of the Earth. The earliest date claimed for life on Earth so far is 4.28 Gyr ago, based on the discovery of microfossils in sedimentary rocks believed to be associated with a seafloor hydrothermal vent (Dodd et al., 2017). This would place the emergence of unicellular life on Earth at a time almost simultaneous with the end of the Hadean epoch 4 Gyr ago, and the end of the late, heavy bombardment, a period of intense asteroid and comet impact activity, signs of which are found throughout the inner Solar System (Bottke & Norman, 2017). An early date for the emergence of life is supported by the results of isotopic analysis of graphite inclusions in a 4.1 Gyr old zircon from Western Australia (Bell et al., 2015).

The discovery of evidence for life on Earth over 4 Gyr ago is rather surprising since the young Earth was a truly hostile environment, with a largely molten surface. If our ideas about the young Earth are correct, then it would appear that life developed on Earth almost as soon as it was possible for it to develop. Needless to say, this has significant implications for the search for extraterrestrial life and intelligence since it suggests that life can emerge very rapidly in compatible environments.

3. **Resources Necessary for Life**

Life on Earth, as we commonly see it, is almost entirely dependent on photosynthesis, and thus light form the Sun. One might therefore think that access to sunlight on the surface of a planet, be it on land or in an ocean, is a requirement. However, as we have seen above, one of the potential sites for the first life on Earth is in fact a deep seafloor hydrothermal vent. Such vents still exist in the deep ocean today, and are well known as oases of biological activity in the otherwise barren ocean depths (eg. Lutz & Kennish, 1993). The primary energy source for these rich biological communities is chemosynthetic rather than photosynthetic, with bacteria using the rich chemical mix emerging from the ocean floor as their primary source of energy. There is even evidence that sub-oceanic basalts contain large communities of chemosynthetic bacteria (Lever et al., 2013) raising the possibility that the largest component of the Earth's biosphere actually lies in the rocks that extend kilometres deep beneath the oceans and cover 60% of the surface of the Earth (Edwards et al., 2012).

From this we can conclude that the availability of energy is the common thread, rather than dependence on any specific form of energy, such as sunlight.

The other common factor that seems to apply to all of the biological communities we are aware of on Earth is water, which provides the liquid substrate in which all biological reactions occur. While there have been suggestions that other solvents might be capable of supporting biological processes, such as ammonia, hydrocarbons and other more exotic compounds (eg. Baross et al., 2007; McLendon et al., 2015), there is to date no evidence supporting these suggestions.

We thus conclude that our consideration of possible locations for life in the Solar System must be restricted to bodies where liquid water and an available supply of energy can be found. While this is a conservative assumption, as we shall see many such locations are available.

## 4. Habitable Bodies of the Solar System

Given these assumptions about the requirements for life, we can make some assessments of the size of potential ecosystems in various Solar System objects by looking at the volume of water available. A viable ecosystem in these locations would then arise if there is a source of energy to both melt some fraction of this available water, and power any emergent ecosystem. Table 1 shows a summary of the amount of water available in a number of different Solar System bodies. While a significant fraction of the water in many of these bodies will be frozen into ice, there is abundant evidence that some of these objects (at least Enceladus, Europa, Titan and Ganymede) contain large bodies of liquid water below their icy surfaces (Vance et al., 2018; Beuthe et al., 2016; Hurford et al., 2017; Keane et al., 2016; Nimmo et al., 2016). The discovery of geysers on Enceladus and Europa, ejecting water and other material from a subsurface ocean (Hanson et al., 2006; Drabek-Maunder et al., 2017; Sparks et al., 2016) provides clear confirmation of the presence of liquid water and a source of energy keeping it liquid in at least these two objects. Meanwhile, though it seems highly likely that water flowed on Mars in the distant past and likely survives there today, a full estimate for the amount of water in this traditional target of searches for extraterrestrial life, is still missing.

| Object | Radius (km) | Water Volume (km$^3$) |
|---:|---:|---:|
| Earth | 6370 | 140 x 10$^7$ |
| Enceladus | 252 | 4.4 x 10$^7$ |
| Dione | 561 | 46 x 10$^7$ |
| Europa | 1565 | 280 x 10$^7$ |
| Pluto | 1187 | 430 x 10$^7$ |
| Triton | 1352 | 670 x 10$^7$ |
| Callisto | 2410 | 2440 x 10$^7$ |
| Titan | 2576 | 2830 x 10$^7$ |
| Ganymede | 2631 | 5440 x 10$^7$ |

**Table 1:** A summary of the available volumes of water in a number of Solar System bodies. Note that many of these bodies have available water volumes up to several tens of times that of the water volume of the Earth. They could thus potentially have much larger

biospheres given a source of energy to feed it, and to ensure the water is liquid and not frozen. Uses data from Steve Vance: NASA/JPL-Caltech.

The most surprising result from these numbers is that the Earth, often referred to as the blue planet because of its abundant water, has far less water than many of the icy moons and minor planets in the Solar System. Even tiny Dione, a small moon of Saturn, has 1/3 of the water of the Earth, while Ganymede has forty times as much.

## 5. Discussion and Conclusions

Since we have deemed that water is the essential environment for life, the implication of these numbers is that the icy moons of the outer Solar System can potentially host much larger ecosystems than the one that is present on the Earth. There are certainly other requirements, such as an adequate supply of energy, potentially from radiogenic or tidal heating, but we know such power sources are present from observations of the plumes on Europa and Enceladus.

Secondly, the history of life on Earth now seems to be showing us that life, albeit very simple prokaryotic monocellular life[1], emerged very rapidly once the Earth had become compatible with its existence.

What are the implications of these results for the Fermi Paradox?

If we put the two sets of results together we find the following:

- The rapid emergence of life on Earth suggests that life can emerge wherever there is a compatible environment
- There are large bio-compatible environments inside the icy moons of Solar System gas giants
- The size of the bio-compatible environments inside gas giant moons is much larger, both in sum and, in the case of the large moons like Ganymede and Titan, individually, than the biosphere of Earth

The conclusion of this analysis for our own Solar System is that the interior of the icy moons may be where the bulk of life in the Solar System is to be found. However, this life, intelligent or otherwise, would be locked beneath many kilometres of solid ice, only able to escape into the broader universe through catastrophic geyser eruptions such as those found on Enceladus and Europa.

There are already indications that moons exist around some exoplanets (Kenworthy et al., 2015), and there is no reason to think that moons around gas giant exoplanets should be any different from those seen in the Solar System. There may thus be a very large number of them. Given the discussion here, it could be that the subsurface oceans of gas giant exomoons are in fact the dominant home for life in the Galaxy, with life on Earthlike, terrestrial planets, being the exception rather than the rule. We note that the prospects for identifying more exomoons and for their detailed study in the next few years are very good (Peters & Turner, 2013; Kipping et al., 2009) using data from *Kepler* or the *James Webb Space Telescope*.

---

[1] An alternative solution to the Fermi Paradox is that, while simple, prokaryotic life might be common in the universe, more complex eukaryotic and multicellular life is not. Eukaryotic life emerged on Earth 2.7 Gyr ago, long after the first emergence of life, while multicellular life only emerged 600 Myr ago.

In the context of the Fermi Paradox, we know that species that live in water can evolve to a high level of intelligence – dolphins and octopuses are good examples (though it should be noted that dolphins are evolutionary returnees to the water). However, a liquid environment may be a limiting factor in the development of technology. The dominant concern in this context, though, is the location of these vast watery environments beneath tens or hundreds of kilometres of water ice. While there may be water geysers in some of these, the larger environments of Titan, Callisto and Ganymede in our own Solar System show no signs of escaping water since they are likely capped by an ice layer over 100 km thick. This probably presents an insurmountable obstacle to any intelligence dwelling there even knowing about the outside universe, let alone attempting to communicate with it.

Further exploration of the content of the oceans beneath the icy moons in our own Solar System through new missions like JUICE (JUpiter ICy moons Explorer), ground-based observations looking for the spectral signatures of organic molecules (Drabek-Maunder et al., 2017), or thorough analysis of existing data from past missions like *Cassini* (Postberg et al., 2018), will help determine the validity of the speculations in this paper. We note that the latest results from *Cassini* suggest the presence of complex biochemical molecules in the ocean of Enceladus, though their origin remains unclear.

We are left with the rather chilling prospect that the galaxy may be filled with life, but that any intelligence within it is locked away beneath impenetrable ice barriers, unable to communicate with, or even comprehend the existence of, the universe outside.

**Acknowledgements**

It is a pleasure to thank Gerry Webb and the BIS for the invitation to contribute to the Fermi Paradox discussion meeting and to this special edition. I would also like to thank my PhD students Josh Greenslade and Tai-an Cheng for useful discussions.